\documentstyle[prl,aps,multicol]{revtex}   

\begin{document}
\draft
\preprint{{\bf ETH-TH/98-04}}

\title{Nonlinear Transport through NS Junctions due to Imperfect
  Andreev Reflection}

\author{Gordey B.\ Lesovik$^{a\,}$ and Gianni Blatter$^{b\,}$}

\address{$^{a\,}$Institute of Solid State Physics, 142432
  Chernogolovka, Moscow District, Russia}

\address{$^{b\,}$Theoretische Physik, ETH-H\"onggerberg, CH-8093
  Z\"urich, Switzerland}

\date{today}
\maketitle
\begin{abstract}
  
 {We investigate a normal metal -- superconductor (point) contact in
  the limit where the number of conducting channels in the metallic
  wire is reduced to few channels. As the effective Fermi energy drops
  below the gap energy, a conducting band with a width twice the Fermi
  energy is formed. Depending on the mode of operation, the conduction
  band can be further squeezed, leading to various non-linear effects
  in the current-voltage characteristics such as current saturation, a
  N-shaped negative differential resistance, bistability, and
  hysteresis.}


\end{abstract}

\begin{multicols}{2}          
\narrowtext

Coherent transport phenomena in micro-structured
normal--superconductor (NS) systems have recently attracted a lot of
interest \cite{general}. The transport across a (NS) boundary is
governed by the phenomenon of Andreev reflection \cite{Andreev}: An
electron incident from the normal metal on the NS junction with an
excitation energy $|\varepsilon|$ below the superconducting gap
$\Delta$ cannot penetrate into the bulk superconductor (we measure the
excitation energy $\varepsilon$ of the electron with respect to the
chemical potential $\mu$ in the superconductor). Nevertheless, subgap
transport across the junction is possible via the process of Andreev
reflection, where the electron incident on the boundary is accompanied
by a (coherent reflected) hole, producing effectively a state with two
incoming electrons which convert into a Cooper pair upon entering the
superconductor. For an ideal NS boundary, such a process leads to a
conductance $G = 4e^2/h$ per channel \cite{BTK}, twice higher than the
maximally possible normal one. If the transparency $T$ of the boundary
is smaller than unity, the NS linear conductance decreases as $T^2$ at
small $T \ll 1$ \cite {BTK}. New effects appear in the finite-bias or
finite-temperature conductance when the transmission of electrons and
holes differs significantly as a consequence of their different
longitudinal kinetic energies \cite{general,Dz,Wendin,LFB}. In this
letter we show how Andreev scattering, combined with specific
conditions for the propagation of electrons and holes, leads to the
formation of a subgap conduction band with a width which strongly
depends on the bias voltage, leading to new transport characteristics
exhibiting a negative differential resistance, bistable, and
hysteretic effects.

To fix ideas, consider a single channel normal metal point contact to
a bulk superconductor, see Fig.\ 1(a). Here we have in mind
geometries such as quantum point contacts realized in heterostructures
\cite{Yaps,WvW} or via manipulations with a scanning
tunneling microscope \cite{STM}.  Given the chemical potential $\mu$,
we define the longitudinal chemical potential $\mu_x = \mu -
\varepsilon_\perp$, where $\varepsilon_\perp$ denotes the transverse
energy of quantization in the normal channel. Here, we are interested
in situations where the longitudinal kinetic energy $K_{e,h}$ at the
Fermi surface is small, such that the condition $\mu_x < \Delta$ is
realized. In this case, electrons with excitation energies
$\varepsilon = E - \mu = K_e - \mu_x > 0$ can propagate through the
contact, whereas the corresponding hole state with the same excitation
energy can propagate only if its kinetic energy $K_{h} = \mu_x -
\varepsilon$ remains positive, i.e., $\varepsilon < \mu_x$ (in this
simplified discussion we assume that the conducting normal channel is
long enough to generate transmission probabilities 0 or 1 only).

An electron incident on the superconductor defines the Andreev state
(in the normal single channel region)
\begin{equation}
  \Psi_\varepsilon (x) = {1 \choose 0} \frac{e^{ik_+ x}}{\sqrt{v_+}} + 
{r_{ee} \choose 0} \frac{e^{-ik_+ x}}{\sqrt{v_+}} + {0 \choose r_{eh}} 
\frac{e^{ik_- x}}{\sqrt{v_-}}, 
\label{AndreevstateN}
\end{equation}
with $k_\pm = \sqrt{2m(\mu_x \pm \varepsilon)}/\hbar$ and $v_\pm =
\hbar k_\pm/m$ [the above states are normalized to carry unit particle
flux, with a normalization $\langle \Psi_\varepsilon,
\Psi_{\varepsilon'} \rangle = 2 \pi \hbar \delta(\varepsilon -
\varepsilon')$, implying that $|r_{ee}|^2 = 1 - |r_{eh}|^2$].
Following the above discussion, the quenching of the hole state for
energies $\varepsilon > \mu_x$, combined with the restriction in the
allowed energies for incident electron states $\varepsilon > - \mu_x$,
leads to the formation of a conducting band of width $2\mu_x$, see
Fig.\ 1(b) and the inset of Fig.\ 2.  Within this band, incident
electrons are (nearly) perfectly reflected into holes, whereas
electrons with energies above this band ($\varepsilon > \mu_x$) are
reflected as electrons and do not carry current (electrons with
energies ($\varepsilon < -\mu_x$) do not enter the normal channel at
all).

In the simplest formulation of the problem we consider a single
channel NS junction. The Andreev states are found by solving the
Bogoliubov-de Gennes equations
\begin{equation}
  \left( \begin{array}{cc}\!-\frac{\hbar^2 \partial_x^2}{2m} - \mu_x &
  \Delta(x)\! \\ \! \Delta^*(x) & \frac{\hbar^2 \partial_x^2}{2m} +
  \mu_x\! \end{array}\right)\left(\begin{array}{c}\! u_\varepsilon(x)
  \\ v_\varepsilon(x) \!\end{array} \right) = \varepsilon \, \left(
  \begin{array}{c}\! u_\varepsilon(x)\! \\ \! v_\varepsilon(x)\! 
  \end{array} \right)
\label{BdeG}
\end{equation}
with the gap function $\Delta(x) = \Delta \Theta(x)$, using the Ansatz
(\ref{AndreevstateN}) in the normal region and
\begin{equation}
  \Psi_\varepsilon (x) = t_e {1 \choose \gamma} \frac{e^{(ip-q)x}}
  {\sqrt{v}}+t_h{1 \choose \gamma^\ast}\frac{e^{(-ip-q)x}}{\sqrt{v}}
\label{AndreevstateS}
\end{equation}
in the superconducting region $x>0$, with $p^2-q^2 = 2 m \mu_x/\hbar^2$,
$pq = m\sqrt{\Delta^2 -\varepsilon^2}/\hbar^2$, $\gamma = (\varepsilon
-i\sqrt{\Delta^2 - \varepsilon^2})/\Delta$, and the normalizing
velocity $v = \hbar p/m$. Solving for the transmission and reflection
coefficients we obtain the spectral conductance $G{\rm
  \scriptscriptstyle NS} (|\varepsilon| < \Delta) \approx (8e^2/h)
\Theta(\mu_x - |\varepsilon|) \sqrt{\mu_x^2-\varepsilon^2}/\Delta$,
valid in the limit $\mu_x \ll \Delta$ (see inset of Fig.\ 2;
note that the Andreev approximation is not valid in the present case).
Assuming a rigid band, implying that the applied bias drops at the
boundary of the reservoir to the normal lead, the finite conduction
band produces a current-voltage ($I$-$V$) characteristics $I =
\int^{eV} d\varepsilon \, G{\rm \scriptscriptstyle NS}/e$ which
saturates at a bias $eV = \mu_x$ (we assume a negative bias $V < 0$,
hence $eV > 0$).  This then is the simplest example where the
quenching of the back-propagating holes limits the width of the
conducting band and entails a non-trivial saturation phenomenon in the
transport characteristics of the NS junction. It is in contrast to the
normal point contact, where an increasing bias $eV$ opens up the
channels, see Refs.\ \cite{GK} (note that a non-trivial structure in
the $I$-$V$ characteristics can occur in a normal point contact as
well, though at much higher voltages $eV > \mu$, see \cite{Pepper}).

The above 1D example cannot be trivially applied to a physical
situation: strictly speaking, we neither have a superconducting
instability nor a normal Fermi liquid in a single channel 1D system.
Let us then discuss the more realistic single channel NS point contact
with a geometry as sketched in Fig.\ 1(a), see also \cite{Yaps}: The
1D normal wire is adiabatically connecting the normal reservoir with
the bulk superconductor. The smooth form of the wire guarantees the
appropriate matching of the wavefunctions in the three device
segments, reservoir, 1D wire, and bulk superconductor. An imperfect
matching leads to a normal reflection of the incident particle at the
NS boundary and thus reduces the subgap transport with its interesting
new features (below we will discuss the consequences of normal
reflection for the device characteristics in more detail). The main
properties of this geometry are the following: i) the confining
potential of the wire produces a narrow conduction band connecting the
reservoir with the NS boundary, ii) the 1D wire is long enough to
guarantee a sharp onset of the transmission (but short enough to let
us ignore strong interaction effects due to the one-dimensionality),
iii) the wide NS contact helps the proper matching of the
wavefunctions, and iv) the large chemical potential in the bulk
superconductor allows us to adopt the Andreev approximation. The
functionality of this device resembles that of the idealized structure
above: whereas in the 1D wire the back-propagation of the hole was
limited by the bottom of the conduction band, in the present situation
holes reflected from the NS boundary with minimal kinetic energy
$\mu_x-\varepsilon < 0$ have to tunnel through the effective potential
due to the transverse quantization in the wire, see Fig.\ 1(c).

As we can apply the Andreev approximation for the geometry of Fig.\ 
1(a), the determination of the conductance is trivial, $G_{\rm
  \scriptscriptstyle NS} (\varepsilon) = (4 e^2/h) \Theta(\mu_x -
|\varepsilon|)$, and the corresponding $I$-$V$ characteristics for the
rigid band model follows from simple integration, see Fig.\ 2.

So far, the determination of the current-voltage characteristics has
been based on a rigid band model, where the voltage drop in the device
occurs at the boundary to the normal reservoir. A more accurate
calculation of the transport current $I(V)$ involves a self-consistent
determination of the charge $\rho(x)$ and the electric potential $e
\varphi(x)$ in the wire, given an applied bias $V$, see, e.g., Ref.\ 
\cite{Larkin}. Here we refrain from such a calculation, but rather
discuss two interesting limiting cases illustrating the potential
features of such a device.

The first case we wish to analyze is the gated wire, where a top gate
placed over the wire modifies its potential $e \delta \varphi \equiv
U(V,V_g)$. In the simplest case, considered by Brown {\it et al.}
\cite{Pepper}, the gate potential $V_g$ follows the applied bias,
$\delta V_g = V$. Assuming further that the wire potential is slaved
to the gate, the band bottom in the wire is lifted by $U = eV$,
implying that backpropagating holes and low energy incident electrons
are cutoff at $|\varepsilon| = \pm (\mu_x -eV)$ rather than $\pm
\mu_x$. Within the Andreev approximation the spectral conductance
$G_{\rm \scriptscriptstyle NS}$ is narrowed down to the interval
$|\varepsilon| < \mu_x - eV$ and takes the form
\begin{equation}
  G_{\rm \scriptscriptstyle NS} (\varepsilon, U(V)) = 
  \frac{4 e^2}{h} \Theta[(\mu_x - U(V)) - |\varepsilon|].
\label{G}
\end{equation} 
A simple integration produces the $I$-$V$ characteristics
\begin{equation}
  I(V< \Delta/e) = \frac{1}{e}\int_0^{eV} d \varepsilon 
  G_{\rm \scriptscriptstyle NS}(\varepsilon, U(V))
\label{current-voltage}
\end{equation}
exhibiting a negative differential resistance (NDR) re\-gime within the
bias interval $\mu_x/2 < eV < \mu_x$, see Fig.\ 2: Increasing the
applied biased $eV$ up to $\mu_x/2$, additional current carrying
states are occupied and the transport current $I$ increases. Going
beyond the value $\mu_x/2$, the rising bottom of the band quenches the
back propagation of the holes and fewer states are available, until at
$eV = \mu_x$ all current carrying states are blocked.

The above NDR phenomenon is further accentuated in a device where the
charge of the wire rather than its potential is fixed --- this is the
second limiting case we wish to study here. The contribution to the
charge density of an individual channel averaged over the wire cross
section is given by
\begin{equation}
  \!\!\rho(x) = 2e\sum_k \{f_{\rm \scriptscriptstyle V}(\varepsilon_k) 
  |u_k (x)|^2 + [1\! -\! f_{\rm \scriptscriptstyle V}(\varepsilon_k)] 
  |v_k (x)|^2\},
\label{density}
\end{equation}
with $f_{\rm \scriptscriptstyle V}(\varepsilon)$ the (bias dependent)
distribution function for the Bogoliubov quasi-particles.  We evaluate
the density in the middle of the wire and allow for a non-zero
potential shift $U$. Using the normalization introduced in
(\ref{AndreevstateN}) we arrive at the form
\begin{equation}
  \rho = \frac{e}{\pi\hbar} \int \!\! d \varepsilon 
  \left[ \frac{1+|r_{ee}|^2}{v_+(\varepsilon)} 
  f_{\rm \scriptscriptstyle V}(\varepsilon) + 
  \frac{|r_{eh}|^2}{v_-(\varepsilon)} 
  [1\! -\! f_{\rm \scriptscriptstyle V}(\varepsilon)]
\right]\! ,
\label{density2}
\end{equation}
with $v_\pm = \sqrt{2(\mu_x-U\pm\varepsilon)/m}$ the velocities of the
quasiparticles. For the case of perfect Andreev reflection the above
expression simplifies to [we assume an open channel configuration with
$0<eV<\mu_x-U$; the occupation numbers are determined by those of the
metallic reservoir, $f_{\rm \scriptscriptstyle V}(\varepsilon) =
\Theta(-\varepsilon+eV)$ at zero temperature]
\begin{eqnarray}
  \rho &=& \frac{e}{\pi\hbar} \sqrt{\frac{m}{2}} 
  \left(\int_{-\mu_x+U}^{eV}d\varepsilon
  \frac{1}{\sqrt{\mu_x-U+\varepsilon}} \nonumber \right.\\
  &+& \left.\int_{eV}^{\mu_x-U}d\varepsilon 
  \frac{1} {\sqrt{\mu_x-U-\varepsilon}} \right).
\label{density3}
\end{eqnarray}
Requiring that the charge difference $\delta\rho=\rho(eV,U)-
\rho(0,0)$ vanishes at any applied bias $V$ leads to the condition
($k_{{\rm \scriptscriptstyle F},x} = \sqrt{2m\mu_x}/\hbar$)
\begin{eqnarray*}
  \delta\rho = \frac{ek_{{\rm \scriptscriptstyle F},x}}{\pi} 
  \left[\sqrt{1-\frac{U\! -\! eV}{\mu_x}}+\sqrt{1-\frac{U\! +\! eV}
  {\mu_x}}-2\right]=0,
\end{eqnarray*}
determining the potential shift $U(V)$ in the wire. Solving for $U$ we
obtain the result
\begin{equation}
  U(V) = -\frac{(eV)^2}{4 \mu_x},\quad 0<eV<\mu_x-U(V).
\label{UV1}
\end{equation}
The negative shift in the wire potential seems quite puzzling at first
sight, but can be easily understood in terms of the reduced group
velocity of the back-propagating holes.  A second solution is found
for a positive shift $U>\mu_x$, where the Andreev scattering is
quenched and all incident electrons are reflected back as electrons
($|r_{ee}|^2 = 1$). With $\delta\rho = (2ek_{{\rm \scriptscriptstyle
    F},x}/\pi) [\sqrt{1-(U-eV)/\mu_x}-1]=0$ we find the shift
\begin{equation}
  U(V) = eV,\quad \mu_x-U(V)<0<eV.
\label{UV2}
\end{equation}
Finally, a regime with partially quenched Andreev scattering is found
in the interval $0<\mu_x-U<eV$, with $U(V)$ determined by the
relation $\delta\rho = (ek_{{\rm \scriptscriptstyle F},x}/\pi)$ 
$[2\sqrt{1-(U-eV)/\mu_x}-\sqrt{2(1-U/\mu_x)}-2]=0$ and the result
\begin{equation}
  U = 2eV - 5\mu_x+4\sqrt{\mu_x(2\mu_x-eV)}.
\label{UV3}
\end{equation}
The internal potential shift $U$ versus applied bias $V$ is shown in
Fig.\ 3 (thick solid line).  The three branches of $U(V)$ exhibiting
completely open, partially quenched, and entirely quenched hole
propagation in the wire arrange to define a typical bistable
configuration of the wire within the bias interval $\mu_x < eV <
2\mu_x$. The lower branch with the open channel terminates at $eV =
2\mu_x$ and the system has to jump to the state where the
backpropagation of holes is quenched. Physically, the jump between the
two branches corresponds to a rearrangement of the potential drop in
the device: At small bias (lower branch) the applied bias drops on the
left side of the channel, towards the reservoir. At high bias (upper
branch) the potential drops on the superconductor side, producing the
gated situation described above. Translating this behavior of the
internal device bias $U$ to the $I$-$V$ characteristics, see Fig.\ 2,
we find a jump from the `open' current carrying state at low bias to
the `closed' gated state at high bias as the applied bias $eV$ grows
beyond the band width $2 \mu_x$, thereby producing a characteristics
with a N-shaped negative differential resistance.  At voltages $eV >
\Delta$ a finite conductance is restored.  Note that the $I$-$V$
characteristics is not symmetric: for a positive applied bias $V>0$
($eV<0$) the conduction band stays open up to the bias $eV =-2\mu_x$,
where the wire potential aligns with the potential in the reservoir,
$U-\mu_x=eV$, see (\ref{UV1}). Increasing the bias further, the
current saturates (similarly to the rigid band case) as part of the
incoming electrons are excluded from entering the wire.

The above analysis for the case of ideal Andreev reflection at the NS
boundary can be easily generalized to take a finite normal
reflectivity of the barrier into account. It is convenient to
characterize the junction through its normal state properties: Given
the reflection coefficient $R$ for electrons entering the 1D-wire, the
parameter $|r_{ee}|^2$ switches between the values $|r_{ee}|^2 =
4R/(1+R)^2$ (`open' channel) and $|r_{ee}|^2 = 0$ (`closed' channel)
(see Refs.\ \cite{Been,LFB}; we assume that $\partial_\varepsilon
R(\varepsilon) \approx 0$ and $\varepsilon \ll \Delta$, allowing us to
ignore the energy dependence in $|r_{ee}|^2$).  A finite value of $R$
then leads to a smoothing of the potential-bias relation and the
current-voltage characteristics, see Figs.\ 2 and 3. As $R$ approaches
unity the fixed-charge characteristics approaches the result of the
gated wire.

Above we have concentrated on the single channel limit, where the
non-linear effects leading to the NDR phenomenon are most pronounced.
Going over to a multi-channel geometry, our analysis can be carried
over to the grazing incidence trajectories \cite{Dz} with the
modification $\mu_x \rightarrow \mu_{x,n} = \mu -
\varepsilon_{\perp,n}$, where $\varepsilon_{\perp,n} = \hbar^2
K_n^2/2m$ is the transverse energy of the $n$-th channel. The
interesting structure obtained in the single channel case (see Fig.\ 
2) then survives for the states with an effective chemical potential
$\mu_{x,n} < \Delta$.  The maximal number of channels to be saturated
or cut off in this fashion below a bias $eV \sim \Delta$ is of order
$\delta n \sim (\Delta/\mu) n$, where $n \gg 1$ is the total number of
channels. In a bulk system with a planar NS boundary these states
correspond to grazing incidence trajectories with angles $\vartheta <
\sqrt{\Delta/\mu}$.

In conclusion, the different propagation conditions for electrons and
holes in NS junctions produce a rich variety of phenomena.  Both, the
zero- and finite-bias anomalies \cite{general} in dirty NS junctions
can be understood in such terms \cite{LFB}. Here, we have shown how
the coherent electron-hole transport in a one/few channel system may
lead to strong non-linearities in the device characteristics,
resulting in an N-shaped NDR $I$-$V$ curve in its most extreme
variant. In conventional semiconductor devices this kind of
instability leads to the formation of domain walls, e.g., the Gunn
effect. In the present case where the transport is non-local and 
coherent we can expect a device operation more similar to that
of a double-barrier resonant-tunneling structure \cite{Goldman}.

We thank Alban Fauch\`ere, Christian Glattli, Ivan Larkin, and Leonid
Levitov for stimulating discussions and the Swiss National Foundation
for financial support.

\vspace{-1.0truecm}

\begin{figure}
  \makebox[1.75in]{\rule[1.125in]{0in}{1.125in}}
  \includegraphics{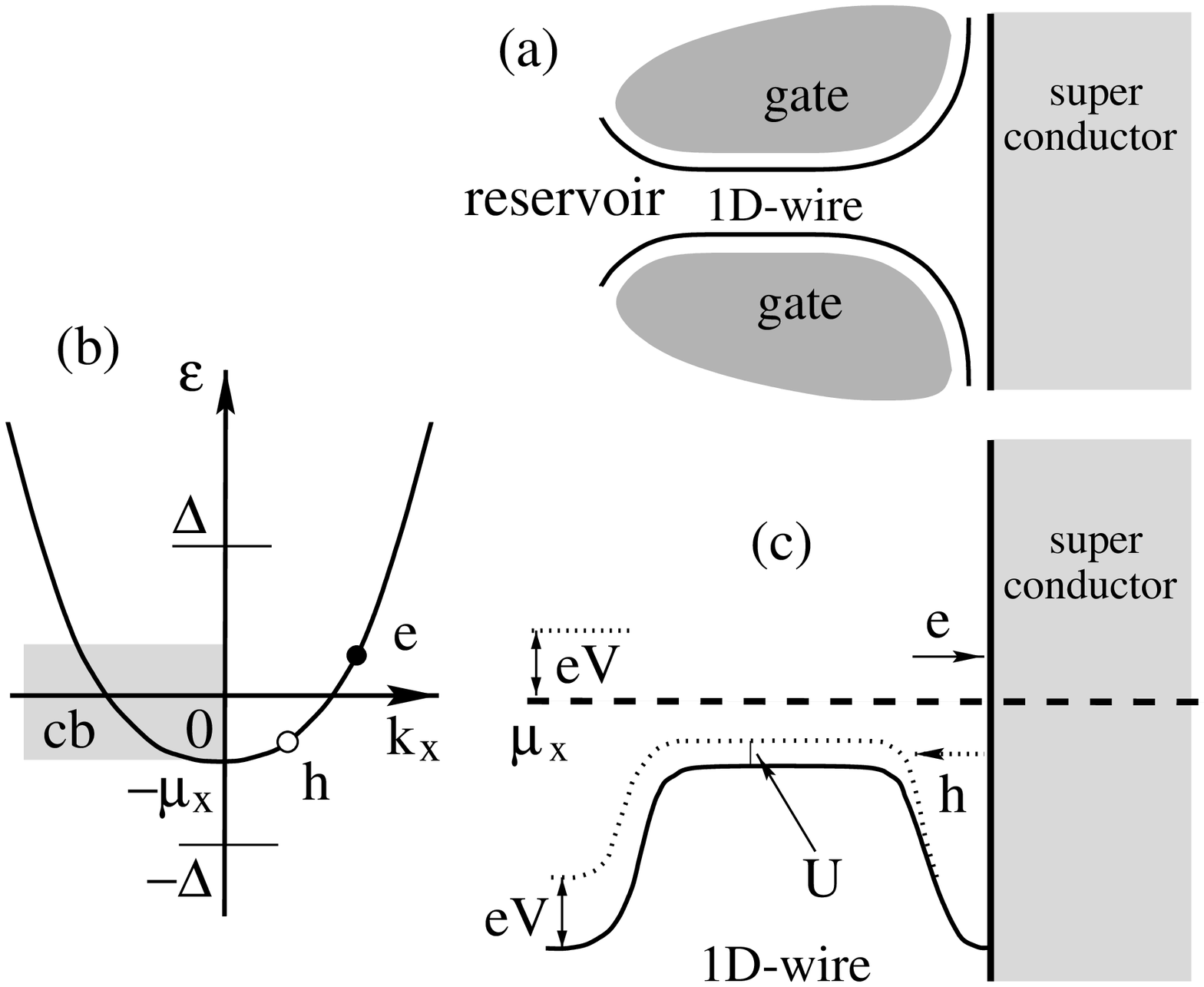} 
\vskip 1.0truecm 
\caption{(a) Geometry of the NS point contact with the 1D normal wire 
  adiabatically connected to the normal reservoir on the left and the
  bulk superconductor on the right, see Ref.\ [8]. (b) Dispersion
  relation in the normal wire. Note the formation of a conducting band
  (cb) of width $2 \mu_x$. (c) Energy diagram for the wire sketched in
  (a). A change in bias $eV$ induces a shift $U$ in the wire
  potential, which in turn may lead to a reflection of the
  back-propagating hole. The electron then is normal-reflected from
  the NS boundary and does not contribute to the current.}
\label{fig:1}
\end{figure}

\begin{figure}
  \makebox[1.75in]{\rule[1.125in]{0in}{1.125in}}
  \includegraphics{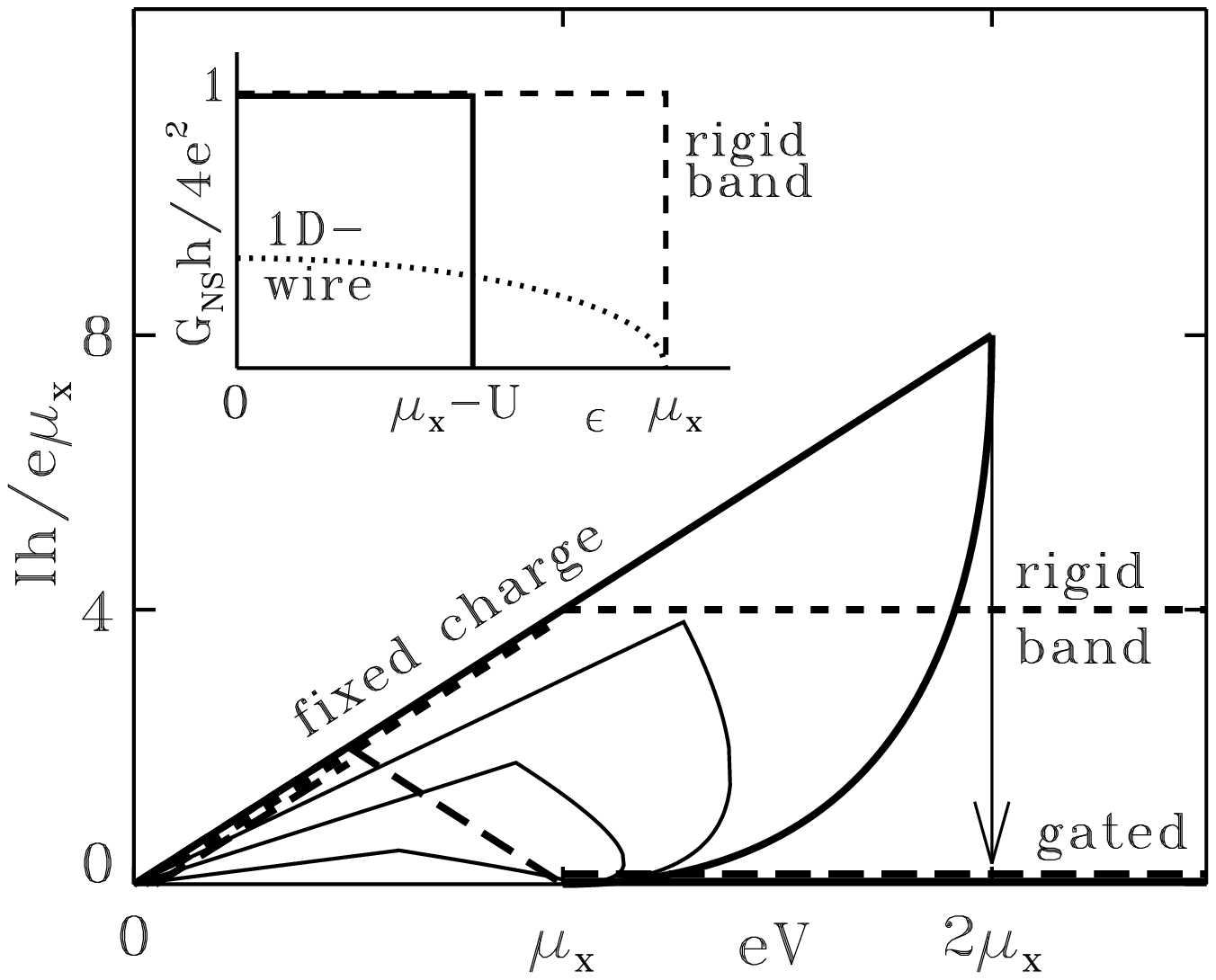} 
\vskip 0.0truecm 
\caption{Current-voltage characteristics of the NS contact. Keeping 
  the band rigid (dashed line), the current saturates when all
  electrons within the band $2\mu_x$ contribute.  In the gated wire
  (long dashes), the back-propagation of holes is partially inhibited
  when the voltage increases beyond $\mu_x/2$ and completely quenched
  beyond $eV = \mu_x$. In the wire with fixed charge (thick solid
  line) the current switches from the upper to the lower branch,
  producing a pronounced negative differential resistivity. A finite
  reflectivity reduces the instability (fixed charge, thin solid
  lines, $|r_{ee}|^2 = 0.25,~0.5,~0.8$).  Inset: Spectral conductance
  $G_{\rm \scriptscriptstyle NS}$ versus energy $\varepsilon$. For the
  1D NS wire the conductance is suppressed due to imperfect Andreev
  reflection (dotted line). For the adiabatically connected wire of
  Fig.\ 1(a) the Andreev approximation is applicable and $G_{\rm
    \scriptscriptstyle NS}$ reaches is maximal value. The width of the
  conduction band depends on the wire potential $U$ (solid line:
  gated/fixed-charge wire; dashed line: rigid band).}
\label{fig:2}
\end{figure}

\begin{figure}
  \makebox[1.75in]{\rule[1.125in]{0in}{1.125in}}
  \includegraphics{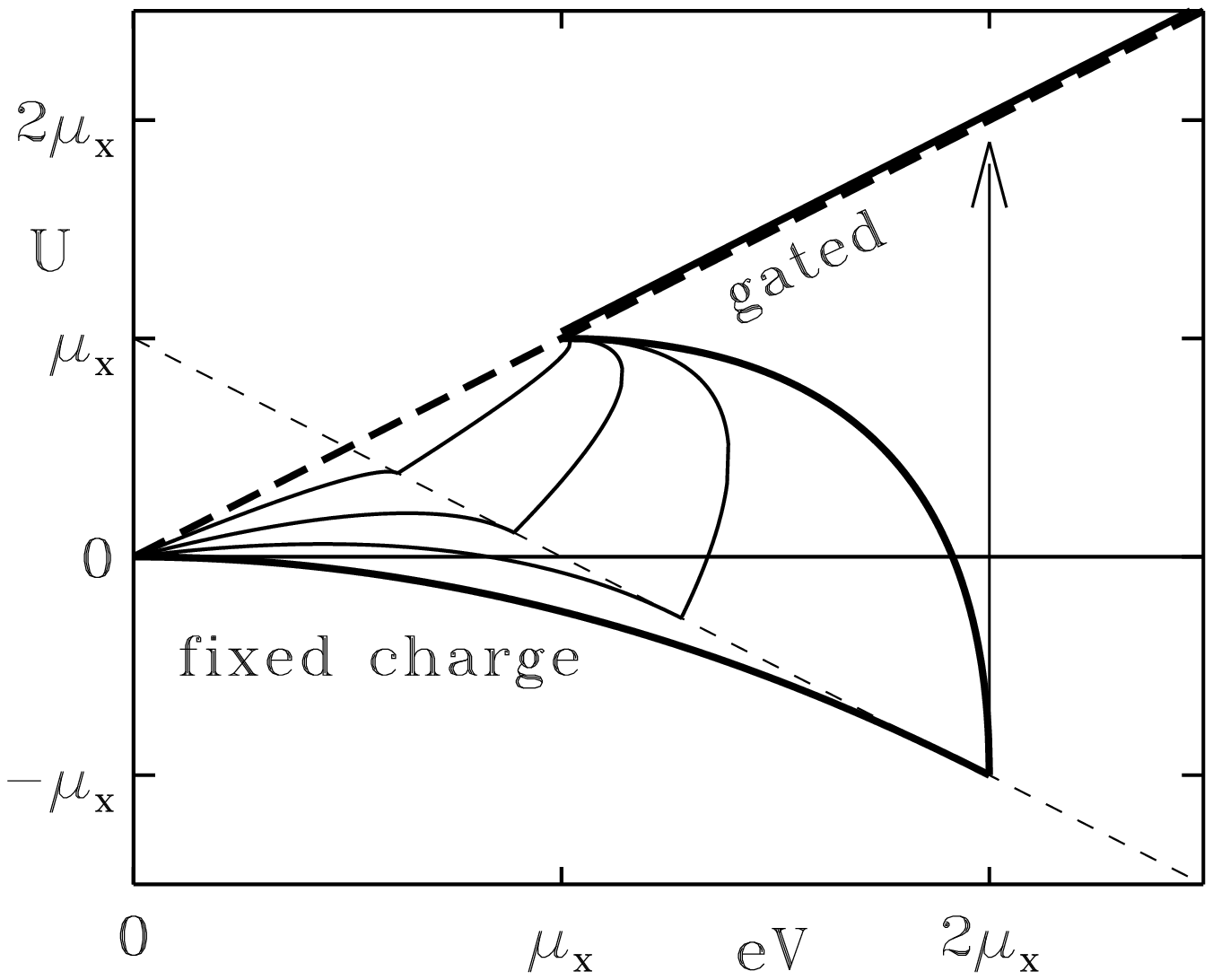} 
\vskip 0.0truecm 
\caption{Potential $U$ within the normal wire versus applied bias for
  the case of a fixed charge and different values of the reflection
  coefficient $r_{ee}$, $r_{ee} = 0$ (thick solid line), $|r_{ee}|^2 =
  0.25,~0.5,~0.8$ (thin solid lines). The lower branch becomes
  unstable at high applied bias and the internal potential $U$ jumps
  to the gated value (long dashes), leading to a NDR in the $I$-$V$
  characteristics of the contact.}
\label{fig:3}
\end{figure}

\end{multicols}          


\vspace{-0.3truecm}

\begin{thebibliography}{99}

\vspace{-1.5truecm}

\bibitem{general} B.\ van Wees and H.\ Takayanagi, in {\it Mesoscopic
    Electron Transport}, eds.\ L.\ Son {\it et al.}, NATO ASI Series
    Vol.\ 345 (Kluwer, 1997), p.\ 469, and references therein;
    C.\ W.\ J.\ Beenakker, in {\it Mesoscopic Quantum
    Physics}, eds. E.\ Akkermans {\it et al.} (Elsevier, 1995).
  
\bibitem{Andreev} A.\ F.\ Andreev, Sov.\ Phys.\ JETP {\bf 19}, 1228
  (1964) [Zh.\ Eksp.\ Teor.\ Fiz.\ {\bf 46}, 1823 (1964)].
  
\bibitem{BTK} G.\ Blonder {\it et al.}, Phys.\ Rev.\ {\bf B 25}, 4515
  (1982).
  
\bibitem{Dz} Yu.\ K.\ Dzhikaev, Sov.\ Phys.\ JETP {\bf 41}, 144 (1975)
  [Zh.\ Eksp.\ Teor.\ Fiz.\ {\bf 68}, 295 (1975)].
  
\bibitem{Wendin} G.\ Wendin and V.\ S.\ Shumeiko, Phys.\ Rev.\ B {\bf
    53}, R6006 (1996).
  
\bibitem{LFB} G.\ B.\ Lesovik {\it et al.}, Phys.\ Rev.\ B {\bf 55},
  3146 (1997).
  
\bibitem{Yaps} H.\ Takayanagi {\it et al.}, Phys.\ Rev.\ Lett.\ {\bf
    75}, 3533 (1995).
  
\bibitem{WvW} D.\ A.\ Wharam {\it et al.}, J.\ Phys.\ C {\bf 21},
  L209 (1988); B.\ J.\ van Wees {\it et al.}, Phys.\ Rev.\ Lett.\ 
  {\bf 60}, 848 (1988).
  
\bibitem{STM} A.\ P.\ Sutton, Curr.\ Opin.\ Solid State Mater.\ Sci.\ 
  {\bf 1}, 827 (1996).
  
\bibitem{GK} L.\ I.\ Glazman and A.\ V. Khaetskii, Europhys.\ Lett.\ 
  {\bf 9}, 263 (1989); N.\ K.\ Patel {\it et al.}, J.\ Phys.\ C {\bf
    2}, 7247 (1990).
  
\bibitem{Pepper} R.\ J.\ Brown {\it et al.}, J.\ Phys.\ C {\bf 1},
  6285 (1989).
  
\bibitem{Larkin} I.\ Larkin and J.\ Davies, Phys.\ Rev. B {\bf 52},
  R5535 (1995).
        
\bibitem{Been} C.\ W.\ J.\ Beenakker, Phys.\ Rev.\ B {\bf 46}, 12841
  (1992).

\bibitem{Goldman} V.\ J.\ Goldman {\it et al.}, Phys.\ Rev.\ Lett.\
  {\bf 58}, 1256 (1987).

\end{thebibliography}
\end{document}